\documentstyle[12pt,twoside,fleqn,espcrc1,epsfig,psfig]{article}

\newcommand{\AmS}{{\protect\the\textfont2
  A\kern-.1667em\lower.5ex\hbox{M}\kern-.125emS}}

\def\J{$J/\psi$}
\def\j{J/\psi}
\def\X{$\chi$}
\def\x{\chi}
\def\P{$\psi'$}

\def\U{$\Upsilon$}

\def\e{\epsilon}

\def\be{\begin{equation}}
\def\ee{\end{equation}}

\def\lsim{\raise0.3ex\hbox{$<$\kern-0.75em\raise-1.1ex\hbox{$\sim$}}}
\def\gsim{\raise0.3ex\hbox{$>$\kern-0.75em\raise-1.1ex\hbox{$\sim$}}}

\def\NP{{ Nucl.\ Phys.\ }}
\def\PL{{ Phys.\ Lett.\ }}
\def\PR{{ Phys.\ Rev.\ }}

\def\PRL{{ Phys.\ Rev.\ Lett.\ }}

\def\ZP{{ Z.\ Phys.\ }}

\title{The Onset of Deconfinement in Nuclear Collisions}
       
\author{H. Satz\address{Fakult\"at f\"ur Physik,
        Universit\"at Bielefeld, \\ 
        Postfach 10 01 31, D-33501 Bielefeld, Germany}%
        \thanks{Different parts of the material
        presented here are based on joint work with S. Fortunato, H.-W. Huang,
        D. Kharzeev, M. Nardi and D. Srivastava.}}
        
\begin{document}
\maketitle

\begin{abstract}
We first consider the origin of colour deconfinement, comparing
in particular spontaneous symmetry breaking and percolation. Next we
specify the onset of deconfinement in nuclear collisions through parton
percolation, and with parameters determined at SPS energy, we give
quantitative predictions for charmonium suppression at RHIC. Finally we
show that the parton cascade model leads to similar
predictions, starting from a microscopic description of the space-time
evolution of the collision.
\end{abstract}

\section{INTRODUCTION}
The interesting new results on the threshold and the further pattern of
anomalous \J~suppression \cite{Cicalo,Kluberg} make it appropriate to
take stock of what we can say about the onset of deconfinement in
nuclear collisions and what we can predict for the forthcoming RHIC
experiments.
The aim of this survey is therefore to formulate the conditions for
deconfinement, use the SPS results to determine the crucial onset
parameters, and then calculate what should be seen at RHIC.

Finite temperature QCD predicts that strongly interacting matter will
undergo a transition at $T=T_c$; at this point, hadronic matter, as the
colour insulator phase of QCD, turns into the colour-conducting
quark-gluon plasma. To throw some light on how and why this happens, we
begin with a study of physical patterns for deconfinement.

\subsection{Spontaneous Symmetry Breaking vs.\ Cluster Percolation}

\medskip

For an infinite bare quark mass, the QCD Lagrangian 
leads to pure colour $SU(N)$ gauge theory. Here the expectation value
of the Polyakov loop $L(T)$ provides an order parameter suitable to
define the confinement-deconfinement phase transition, driven by the
underlying center $Z_N \!\subset\! SU(N)$ symmetry of the Lagrangian
\cite{MS,KPS}.
Confined states share this symmetry, but it is spontaneously broken in
the deconfined phase. Spin systems, in particular the $N$-state Potts
model ($N$=2 is the Ising model), show the same kind of critical
behaviour, based on the spontaneous breaking of a global $Z_N$ symmetry,
with the `magnetisation' $m(T)$ as order parameter. This similarity has
led to the conjecture that the two systems are in fact in the same
universality class of critical behaviour \cite{S-Y}, which is
indeed nicely confirmed in lattice studies \cite{Engels}.

To arrive at an alternative physical definition of deconfinement, we
will make use of the relation between spin and gauge systems. The
critical behaviour of the Ising model can today be characterized either
as
\begin{itemize}
\item{spontaneous magnetisation, i.e., as a disorder/order transition
connected to the spontaneous $Z_2$ symmetry breaking for $T\leq T_c$,
or as}
\item{percolation of clusters of parallel spins joined by suitable
bonds, i.e., in terms of the geometric onset of the new phase.}
\end{itemize}
These two descriptions have been shown to be completely equivalent,
leading to the same critical temperature and the same critical exponents
\cite{C-K,S-A}.

Since the percolation approach to the critical behaviour of spin systems
is fairly new, we briefly sketch its basic features. For simplicity,
consider a two-dimensional square lattice of linear size $R$; on the
$R^2$ sites, we distribute an equal number of up and down spins
according to the Ising Hamiltonian $H_I(T)$. With decreasing temperature
$T$, adjacent sites of parallel spins will begin to form growing
clusters, and for $T \leq T_c$, either the up or the down spins will
percolate, i.e., form an infinite cluster in the thermodynamic limit
$R\to \infty$. We thus can define the percolation strength
\be
P(T) \sim \left(1 - {T \over T_c}\right)^{\beta_p},~~~T \leq T_c,
\label{1.1}
\ee
which measures the probability that an arbitrary site belongs to the
infinite cluster. Since $P(T)=0$ for all $T \geq T_c$ and non-zero for
all $T < T_c$, it constitutes an order parameter for percolation.
Although the critical
temperature $T_c$ for this transition coincides (in two dimensions) with
the Curie temperature, the critical exponents in general do not. To
achieve complete equivalence, including identical critical exponents,
the definition of cluster has to be modified such that percolation
clusters coincide with the correlation clusters obtained in the Ising
model \cite{C-K}. This is achieved by assigning to pairs of
adjacent aligned spins in a geometric cluster an additional bond
correlation, randomly distributed with the density
\be
n_b = 1 - e^{-2J/kT}, \label{1.2}
\ee
where $2J$ just corresponds to the energy required for flipping
an aligned into a non-aligned spin. The modified percolation clusters
then consist of aligned spins which are bond connected. Only for $T=0$
are all aligned spins bonded; for $T >0$, some aligned spins in a
purely geometric cluster are not bonded and hence do not belong to the
modified cluster. This effectively reduces the size of a given cluster
of aligned spins or even cuts it into several modified clusters.
With such a cluster definition, cluster percolation and spontaneous
magnetisation become fully equivalent as critical phenomena.

We now want to extend the percolation definition of critical behaviour
to deconfinement in $SU(2)$ gauge theory, where we consider clusters of
positive (or negative) Polyakov loops $L_i$ instead of up or down spin
clusters; here $i$ denotes the space position of the Polyakov loop.
First studies have been carried out in two space dimensions \cite{FHS},
using
\be
n_b = 1 - e^{-2 \kappa L_iL_j} \label{1.3}
\ee
as bond weight between the (same-sign) Polyakov loops $L_i$ and $L_j$
on adjacent spatial lattice sites $i$ and $j$. The effective coupling
$\kappa$ is known for small temporal lattices \cite{KG}; we therefore
employ for our first study $N_{\tau}=2$. Using finite size scaling
to extrapolate to the thermodynamic limit of large systems, our study
provides two main results:
\begin{itemize}
\item{the percolation coupling (``temperature") coincides with the
deconfinement value obtained using the expectation value of the
Polyakov loop as order parameter; and}
\item{the critical exponents for percolation strength and cluster size
coincide with the Ising values predicted by the Svetitsky-Yaffe
conjecture for $SU(2)$ thermodynamics.}
\end{itemize}
For the specific system studied [$SU(2)$ gauge theory, two space
dimensions, $N_{\tau}=2$], we thus find that deconfinement can be
specified in terms of Polyakov loop percolation \cite{FHS} with as good
or better precision than through the spontaneous symmetry breaking
measured by Polyakov loop expectation values \cite{MS,KPS}. The next
steps in this program are quite clear -- one has to extend $N_{\tau}$ to
arbitrary values and consider also systems in three space dimensions;
then the entire study has to be repeated for $SU(3)$ gauge theory.

One particular advantage of specifying deconfinement in terms of
percolation is that this appears readily generalizable to full QCD
\cite{Satz-P}. In the presence of dynamical quarks, the global
$Z_N$ symmetry of the pure gauge theory Lagrangian is explicitly
broken, so that in full QCD the expectation value of the Polyakov loop
no longer serves as an order parameter for a transition -- it is always
non-vanishing, even though it continues to vary rapidly in a 
\setlength{\unitlength}{1.cm}
\begin{picture}(16,9.4)
  \setlength{\unitlength}{1.cm}
  \put(-0.1,9){
    \begin{minipage}[t]{9.cm}
narrow temperature range. 
In contrast, the percolation strength $P(T)$ remains
an order parameter (for the percolation transition) even for finite
quark mass $m_q < \infty$. This
situation corresponds to the Ising model with non-vanishing external
field $H$; here there is also no more spontaneous symmetry breaking,
with the magnetisation $m(T,H\!\not=\!0) \not= 0$ for all $T$. Nevertheless,
the percolation transition persists for all values of the temperature,
tracing a curve $T_p(H)$ of critical temperature (the so-called
`Kert\'esz line' [8,12]) in the $T-H$ plane. Its counterpart in QCD
defines the deconfinement diagram shown in Fig.\ 1, 
separating the
confined and the deconfined phase by a line of genuine (percolation)
critical behaviour, not just by a rapid cross-over. It is obviously of
great interest to see if this transition occurs at those temperature
values where the expectation value of the Polyakov loop shows its rapid
variation. Moreover, the relation between the percolation transition
and the chiral transition at $m_q=0$ poses an interesting open problem.
  \end{minipage}
  }
\put(9.5,-.2){\psfig{file=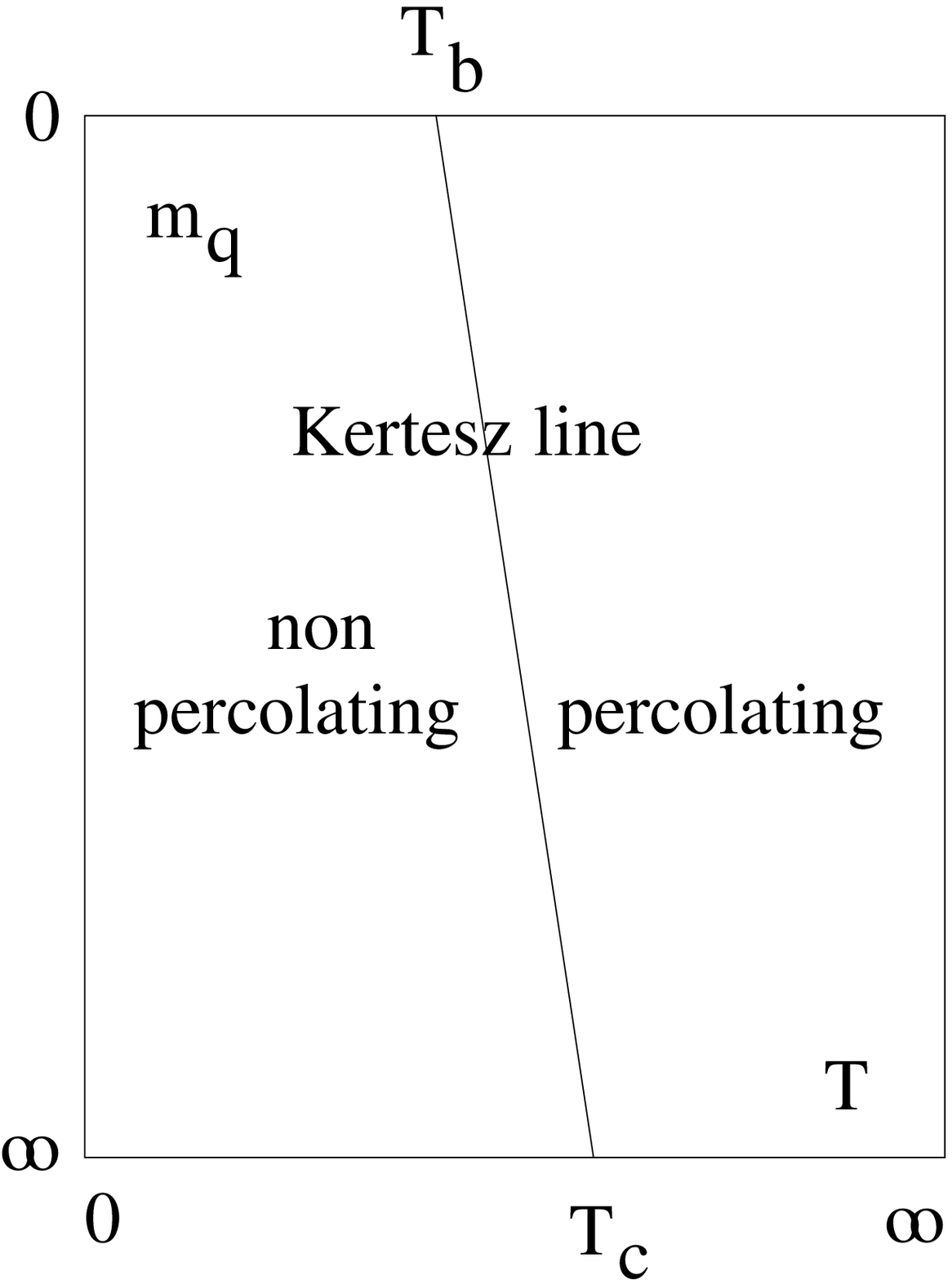,width=6.5cm}}
\put(9.5,-1.2){\mbox{~~Figure 1. The Kert\'esz line of QCD}}
\addtocounter{figure}{1}
\label{fig11}
\end{picture}

\newpage

\subsection{Colour Screening and Quarkonium Suppression}

\medskip

In media so dense that several hadrons overlap, the `other' colour
charges screen the binding forces between the quarks of any given
hadron. As a result, the constituents of the medium become
deconfined, since screening suppresses the long-range confining
potential
\be
\sigma r \to \sigma r \left( {1-e^{-\mu r} \over \mu r} \right),
\label{1.4}
\ee
in a form similar to the Debye screening of the Coulomb potential. The
screening mass $\mu$ is the inverse of the screening radius; it
increases with temperature, so that the range of the potential
decreases, and eventually deconfinement sets in. Finite
temperature lattice QCD determines the critical temperature for
deconfinement; it also indicates that $\mu_c \equiv \mu(T_c)
\sim 0.4 - 0.6$ GeV. With 0.3 - 0.5 fm, the critical colour screening
radius is therefore considerably smaller than the separation between quarks
inside a hadron (about 1.5 - 2 fm).

Colour screening provides a probe of the quark-gluon plasma
\cite{Matsui}. Sufficiently small bound states will not melt
directly at the deconfinement point; they require a shorter
screening radius and hence a hotter quark-gluon plasma to be
dissociated \cite{KMS,KS}. Hence quarkonia can be used to
to measure the temperature of deconfining media.
As example, consider charmonium dissociation. The screening masses
for the dissociation of the \X(3.5) and the \J(3.1) are about 0.35
GeV and 0.70 GeV, respectively \cite{KMS}. The \X~thus melts essentially
at deconfinement, while the \J~requires a somewhat higher temperature.
Similar arguments show that the \U~melts at a very much higher
temperature. For quarkonium states in a uniform medium we thus obtain
a hierarchy of suppression.

Summarizing the first section of this survey, we note
\begin{itemize}
\item{percolation provides a natural specification and a prospective
general order parameter for the onset of deconfinement;}
\item{the dissociation of quarkonia by colour screening signals
deconfinement and serves as thermometer of the deconfined medium.}
\end{itemize}
In the following section, we turn to percolation in nuclear collisions
and to the resulting suppression of \X~and direct \J~production.

\section{PARTON PERCOLATION\cite{KNS}}

In a high energy nuclear collision viewed in the overall center of mass,
the two Lorentz-contracted nuclei quickly pass through each other. After
about 1 fm at the SPS and only about 0.1 fm at RHIC, they have separated
and left behind the partonic medium which is our potential QGP
candidate. Inside this medium, primary nucleon-nucleon collisions have
left probes, such as heavy quark pairs or Drell-Yan dileptons, whose
fate can provide information about the state of the partonic medium.

The average number of produced partons will depend on the number of
nucleon-nucleon collisions and/or the number of participating
(`wounded') nucleons. Up to SPS energies, the number of wounded nucleons
appears to be the main determining factor \cite{Bialas}.
This is understandable in
terms of the dilated and hence large soft parton formation time as seen
in the rest frame of either nucleus. Interference and cancellation
effects of the Landau-Pomeranchuk type thus prevent soft parton
emission at each nucleon-nucleon collision. At higher energies, large
additional contributions can come from hard partons (minijets or jets)
with short formation time, and these will be proportional to the number
of collisions. Even at SPS energy, nucleon stopping and secondary
multiplicities at mid-rapidity seem to increase somewhat with
increasing mass number $A$, and this could be the onset of collision
dependent effects. The $s$-dependence of the mid-rapidity multiplicity
is presumably a combination of the $s$-dependence of the parton
distribution function and the onset of significant hard hadron
production. We shall try to include these dependences here in rather
phenomenological terms, without specifying their origin.

The number of gluons of transverse momentum $k_T$ emitted by a wounded
nucleon can be estimated using the nucleonic gluon distribution function
$g(x)$ from deep inelastic scattering studies,
\be
\left( {dN_g\over dy dk_T^2} \right)_{y=0} \simeq xg(x) f(k_T^2)
\label{2.1}
\ee
where we have assumed factorization of the parton distribution in $x$ 
and $k_T$, and taken  $f(k_T^2)$ as normalized to unity.
At $y=0$, $x\simeq k_T/\sqrt s$, where $\sqrt s$ is the incident collision
energy. Using the MRS-H form of
the gluon distribution function \cite{MRS-H} and integrating the
resulting Eq.\ (\ref{2.1}) over $k_T$ then gives us $(dN/dy)_{y=0}
\simeq 2$ for SPS and 4 for RHIC energy.

Each parton of transverse momentum $k_T$ has an effective transverse
size $r \simeq k_T^{-1}$, and nucleus-nucleus collisions provide many
such partons overlapping in the transverse plane. We want to study their
percolation behaviour. To simplify matters, we shall consider `average'
partons (rather than averaging results with $f(k_T^2)$). From
the transverse momentum dependence of Drell-Yan dilepton production
through quark-antiquark annihilation, or from that of \J~production
through gluon fusion, we know that the effective intrinsic transverse
momentum of partons is $\langle k_T \rangle \simeq 0.75$ GeV/c, leading
to an average transverse parton radius $r \simeq 0.27$ fm. We now
consider the percolation pattern for discs of that size.

\begin{figure}[p]
\vspace*{-0.9cm}
\mbox{
\psfig{file=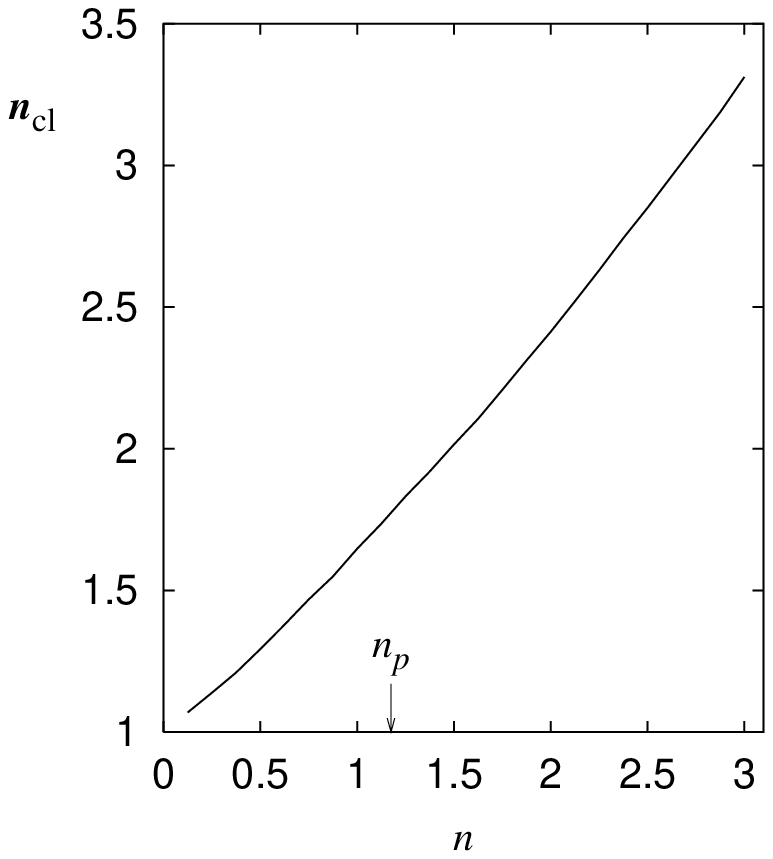,width=8cm}\hskip-.6cm
\psfig{file=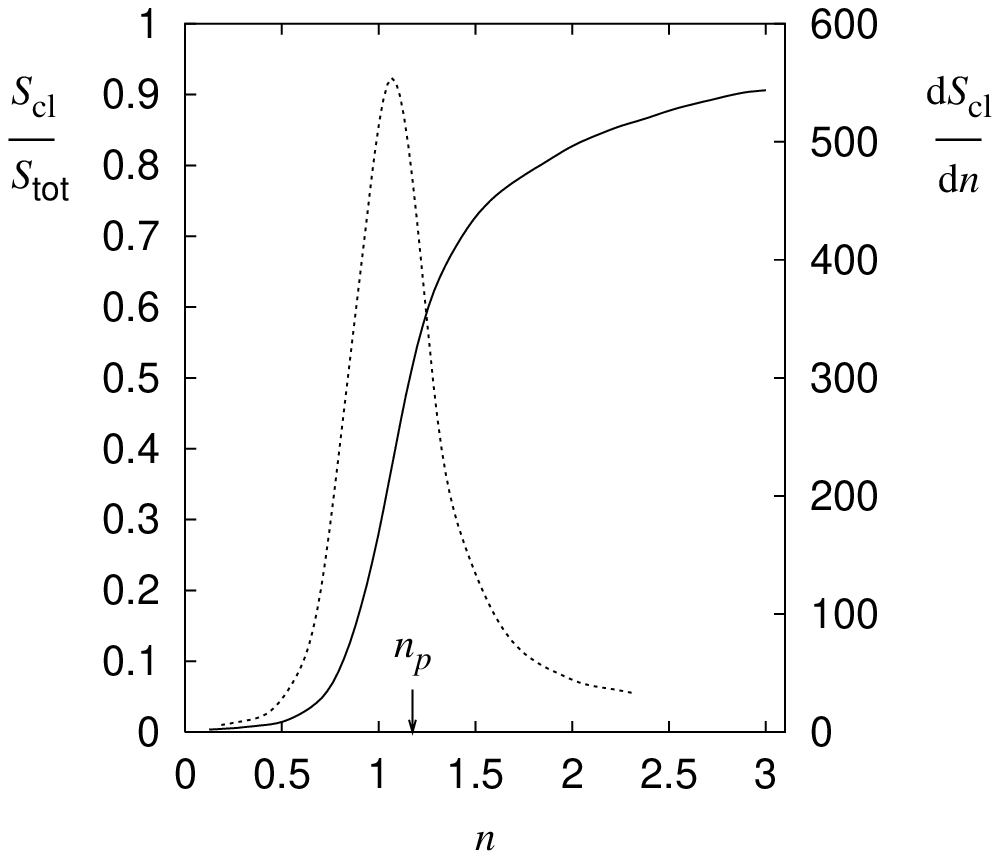,width=8cm}}
\vspace*{-0.8cm}
\caption{Average cluster density $n_{\rm cl}(n)$ (left) and average
fractional cluster size $S_{\rm cl}(n)/S_{\rm total}$ (right) as function of
the overall density $n$ of discs, for $r/R=1/20$; in (b), the derivative
of $S_{\rm cl}(n)/S_{\rm total} $ with respect to $n$ is also shown
(dotted line). The
percolation point in the limit $r/R \to 0$ is indicated by $n_p$.}
\label{fig21}
\vspace*{-0.8cm}
\end{figure}

As a prelude, we assume $N$ discs of radius $r$ to be randomly
distributed on a flat surface of radius $R\!>>\!r$ and study the average
cluster density $n_{\rm cl}(n)$ and the average cluster size $S_{\rm
cl}(n)$ as a function of the overall density $n=N/\pi R^2$. The result
is shown in Fig.\ \ref{fig21}, where we have normalized all quantities to the
inverse disc area $1/\pi r^2$ in order to make them dimensionless. We
note that the derivative of the cluster density peaks sharply at a
certain density; in the `thermodynamic' limit $R \to \infty$, it diverges at
the percolation point $n_p=1.175$ \cite{Alon}.
We see in particular that if we want to reach a certain cluster density
$n_{\rm cl}$, attained at a certain overall $n$, then at this point the
average cluster $S_{\rm cl}(n)$ also has a certain finite size. In other
words, requiring a specific density for the onset of a new phase
implies that this onset occurs for a specific finite fractional size of the
system. The percolation point, for example, is reached when about 50\%
of the surface is covered by discs.

\begin{figure}[p]
  \begin{center}
    \mbox{
      \begin{minipage}{7.5cm}
        \psfig{file=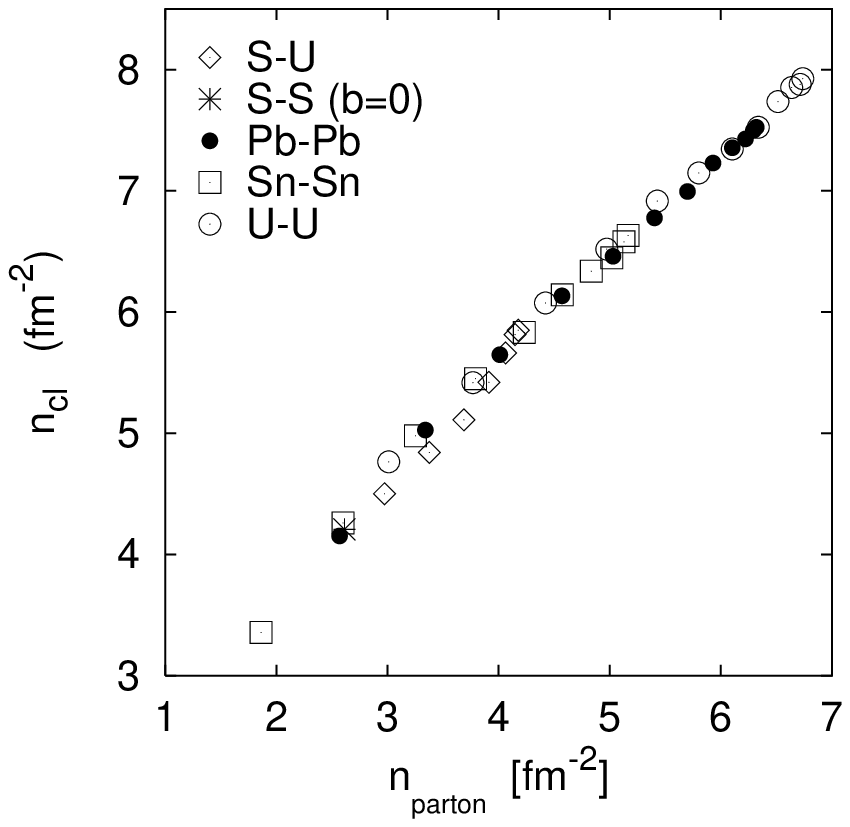,width=7.5cm}
\vspace*{-1.5cm}
        \caption{Cluster density vs.\ parton density for different
          centralities and different $A\!-\!B$ configurations at SPS energy.}
       \label{fig22}
     \end{minipage}
      \hskip.5cm
      \begin{minipage}{7.5cm}
        \psfig{file=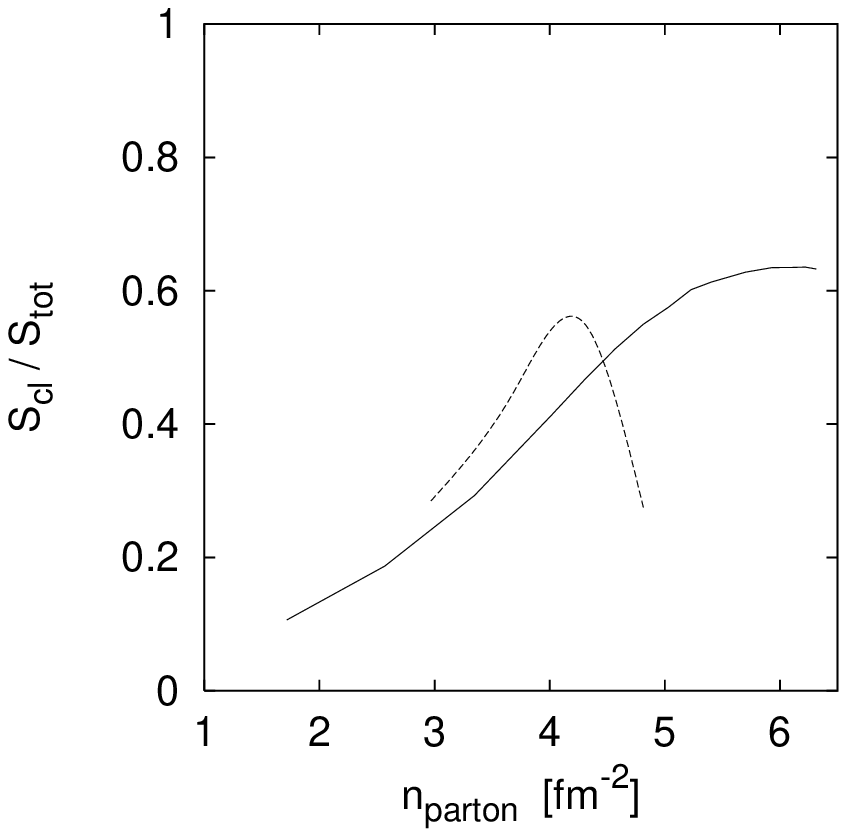,width=7.5cm}
\vspace*{-1.5cm}
        \caption{Fractional cluster size vs.\ parton density, together
          with its derivative, for $Pb\!-\!Pb$ collisions at SPS energy.}
        \label{fig23}   
     \end{minipage}
      }
  \end{center}
\end{figure}

\begin{figure}[h]
  \begin{center}
    \mbox{
      \begin{minipage}{7.5cm}
        \psfig{file=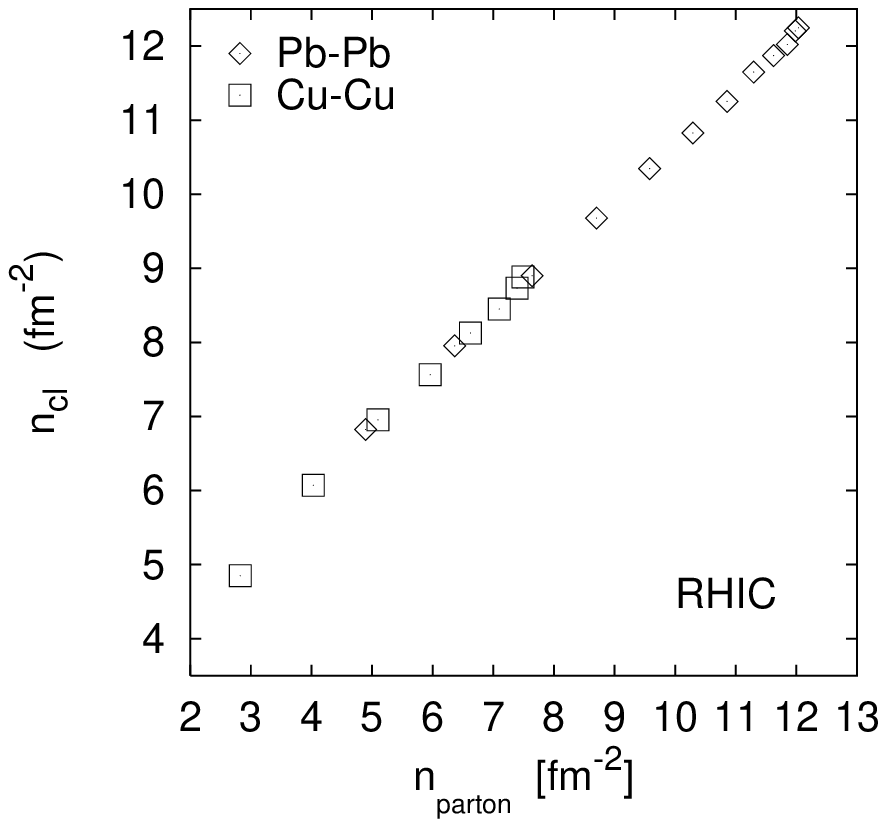,width=7.5cm}
\vspace*{-1.5cm}
         \caption{Cluster density vs.\ parton density for different
          centralities and different $A\!-\!B$ configurations at RHIC energy.}
   \label{fig24}
    \end{minipage}
      \hskip.5cm
      \begin{minipage}{7.5cm}
        \psfig{file=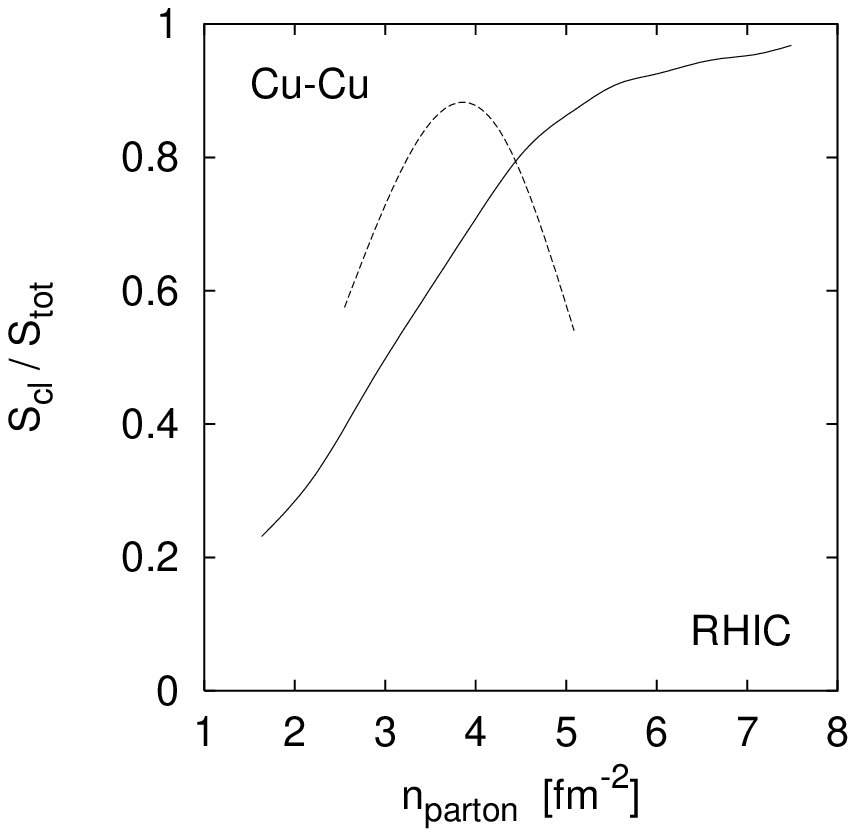,width=7.5cm} 
\vspace*{-1.5cm}
        \caption{Fractional cluster size vs.\ parton density, together
          with its derivative, for $Cu\!-\!Cu$ collisions at RHIC energy.}
      \label{fig25}
 \end{minipage}
      }
\vspace*{-0.5cm}
  \end{center}
\end{figure}

Turning now to nuclear collisions, we have to distribute the partonic
discs not on a flat surface, but instead in the transverse plane
according to the nucleon distribution determined by the profiles of the
colliding nuclei \cite{W-S}. Moreover, we have to allow collisions at
different impact parameters, and we want to consider different $A\!-\!B$
collisions.

In Fig.\ \ref{fig22}, the cluster density $n_{\rm cl}$ is shown as function of
the overall density $n_{\rm parton}$ of `average' partons of radius
$r=0.27$ fm, for different centralities of various $A\!-\!B$
combinations, all at SPS energy, where we take 2 partons per wounded
nucleon. For each $A\!-\!B$, the point at highest $n_{\rm parton}$
corresponds to central (impact parameter $b=0$) collisions, the one at
lowest $n_{\rm parton}$ to the most peripheral collisions possible. We
note that by varying $A\!-\!B$ and centrality we produce an essentially
universal cluster density $n_{\rm cl}(n_{\rm parton})$. Now we assume
that deconfinement occurs at the percolation point, which can be
determined by studying the average fractional cluster size $S_{\rm
cl}/S_{\rm total}$. In Fig.\ \ref{fig23}, we see the result for $Pb\!-\!Pb$
collisions. At the density for which the derivative peaks, $n_{\rm
parton} \simeq 4.2$ fm$^{-2}$, the cluster density reaches its critical
value $n_{\rm cl} \simeq 6$ fm$^{-2}$. The same critical value is found
for $Sn\!-\!Sn$ and $U\!-\!U$ collisions at SPS energy, even though this
value is attained at quite different centrality in the different
configurations ($b \simeq 4.5$ fm for $Sn\!-\!Sn$, 8.5 fm for $Pb\!-\!Pb$ 
and 10 fm for $U\!-\!U$ collisions).

In Fig's.\ \ref{fig24} and \ref{fig25}, we show 
the result of this procedure for RHIC
energy, where we have 4 gluons per wounded nucleon. Again the critical
cluster density at deconfinement is found to be $n_{\rm cl} \simeq 6$
fm$^{-2}$. Here we have to consider $Cu\!-\!Cu$ collisions in order to
study the onset of deconfinement, since for $Pb-Pb$ all possible
centralities are above the threshold.

We thus conclude that the critical cluster density is, as expected, a
universal quantity, independent of the choice of $A\!-\!B$, of
centrality,
and of incident energy. The fractional size of the deconfined cluster at
threshold does, however, depend on the collision configuration: for
$Pb\!-\!Pb$ at SPS, we have $S_{\rm cl}/S_{\rm total} \simeq 0.45$,
while $Cu\!-\!Cu$ at RHIC gives 0.70. In other words, at the higher RHIC
energy, deconfinement sets in for a larger bubble than at the lower SPS
energy.

\begin{figure}[h]
\vspace*{-0mm}
\centerline{\psfig{file=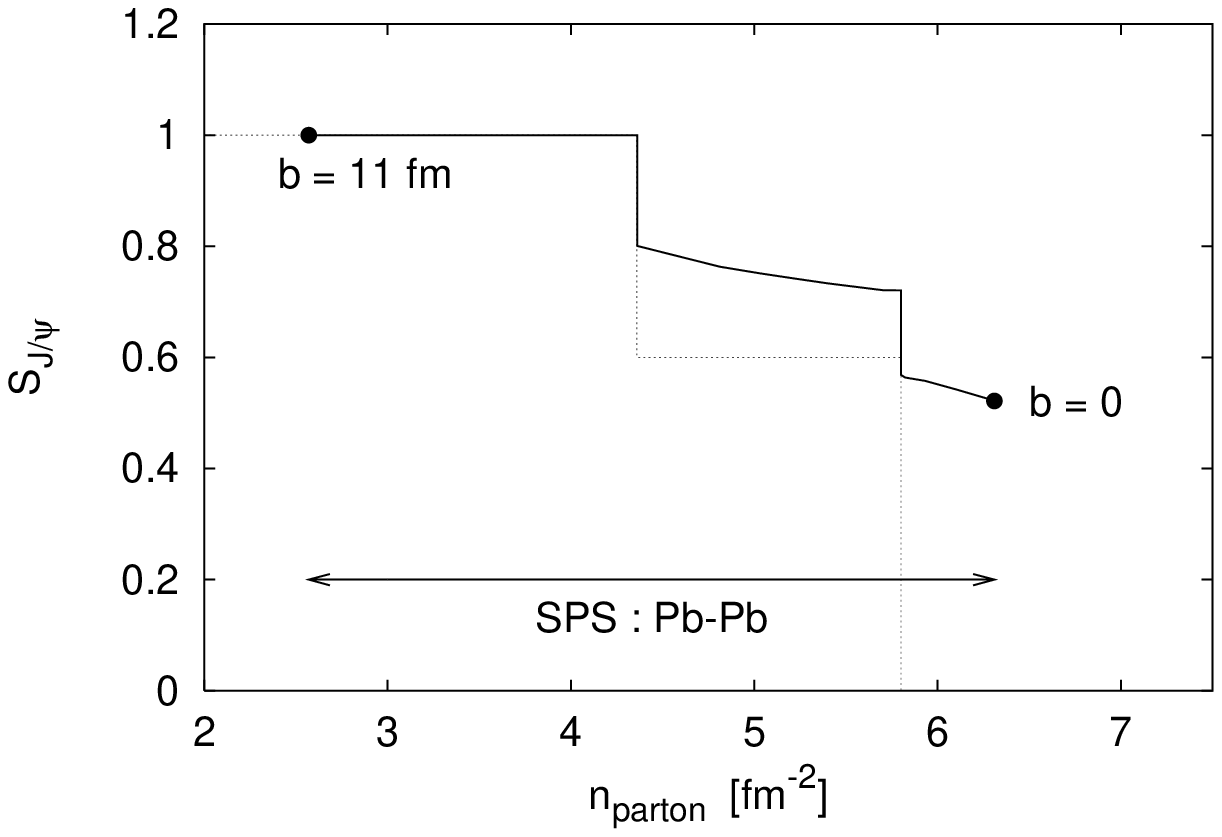,height= 80mm}}
\vspace*{-1.1cm}
\caption{The \J~survival probability as function of the parton density for
$Pb\!-\!Pb$ collisions at SPS energy; the dotted line corresponds to a
medium of uniform parton density, the solid line to collisions with
parton densities determined by the profiles of the colliding nuclei.}
\label{fig26}
\vspace*{-0.5cm}
\end{figure}

\begin{figure}[h]
  \begin{center}
    \mbox{
        \psfig{file=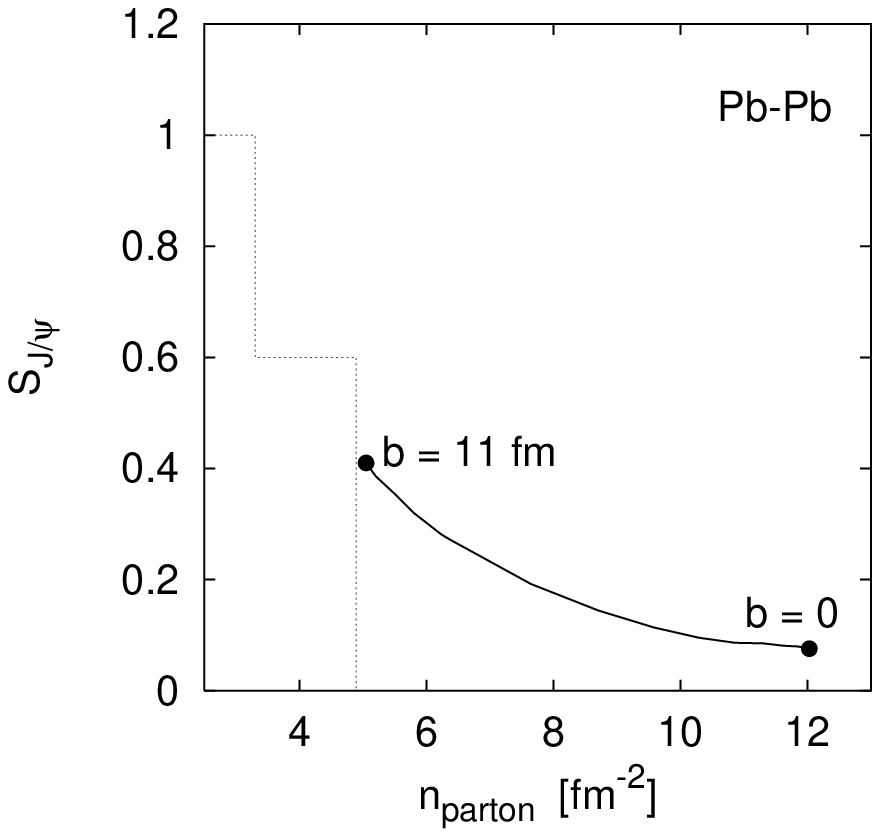,width=7.5cm}
      \hskip.5cm
        \psfig{file=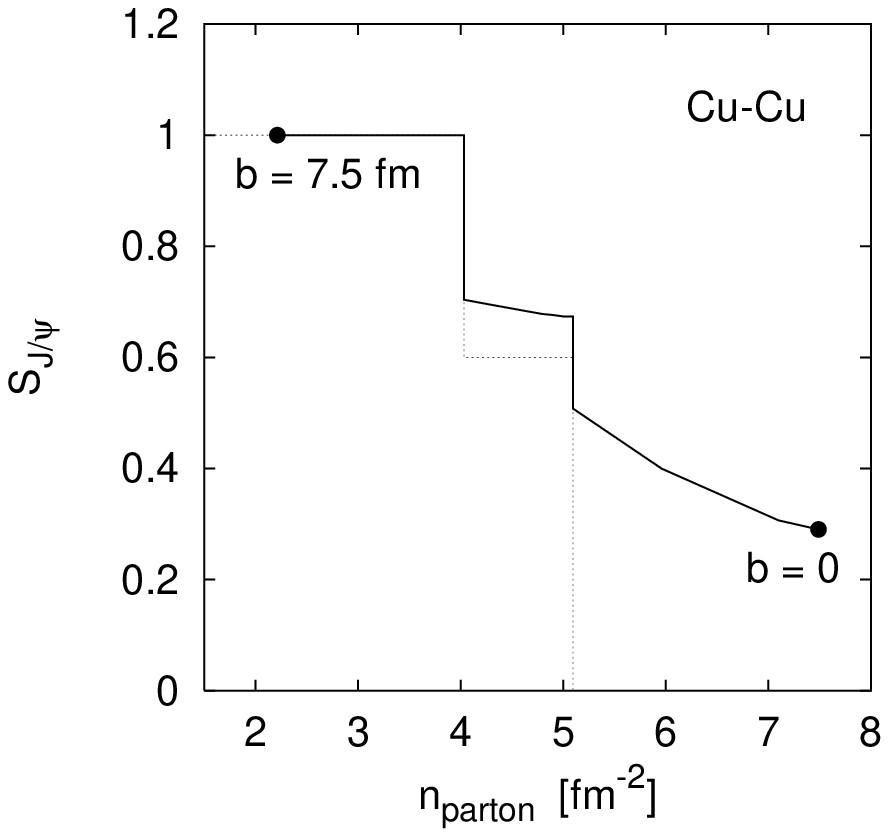,width=7.5cm}
}
\vspace*{-1.2cm}
        \caption{The \J~survival probability as function of the parton 
density for
$Pb\!-\!Pb$ (left) and for $Cu\!-\!Cu$ (right) collisions at RHIC 
energy; the dotted line
corresponds to a medium of uniform parton density, the solid line to
collisions with parton densities determined by the profiles of the
colliding nuclei.}
      \label{fig27}  
  \end{center}
\vspace*{-0.8cm}
\end{figure}

We further note that at SPS energy, $S\!-\!S$ and $S\!-\!U$ are below
the deconfinement threshold even for the most central collisions, while
at RHIC even the most peripheral $Pb\!-\!Pb$ collisions are above the
threshold. To study the threshold, we thus require heavy nuclei
(such as $Pb$) at the SPS, lighter nuclei (such as $Cu$) or lower
incident energy at RHIC.

Having specified the onset of deconfinement through percolation, we now
turn to deconfinement signatures. As noted above, the \X~is dissociated
essentially at deconfinement, so when the cluster density reaches
$n_{\rm cl}^{\rm crit}\simeq 6$ fm$^{-2}$, all \X's which are inside
the percolating cluster disappear. The amount of \X~suppression thus
depends on the fractional cluster size at deconfinement, which is about
45 \% for $Pb-Pb$ at the SPS, about 70 \% for $Cu-Cu$ at RHIC and
effectively 100 \% for $Pb-Pb$ at RHIC. At deconfinement, the
corresponding fractions of \J~production through intermediate \X~states
should thus be suppressed.

As already noted, the dissociation of directly produced \J's requires a
a larger screening mass and thus higher density than available at the
point of deconfinement. On a microscopic level, \J~break-up becomes
possible only for harder than the average gluons at deconfinement. In
principle, the relevant quantities can be determined by lattice QCD
studies; however, this requires computer capabilities which are being
reached only now. For the time, we therefore choose $r_{\j}=0.22$ fm
and $n_{\rm cl}(\j)=7.8$ fm$^{-2}$; this corresponds to a ratio of
transverse dissociation energy densities $\e_T(\j)/\e_T(\x) \simeq
1.6$, in accord in studies combining potential theory and lattice
results \cite{KMS,KS}. With these parameters, one obtains 20\% direct
\J~suppression in $Pb\!-\!Pb$ collisions at SPS energy, once the
required cluster density is reached. The corresponding fractions for
RHIC are 30\% for $Cu\!-\!Cu$ and 35\% for $Pb\!-\!Pb$ collisions.

\begin{figure}[h]
\vspace*{-0mm}
\centerline{\psfig{file=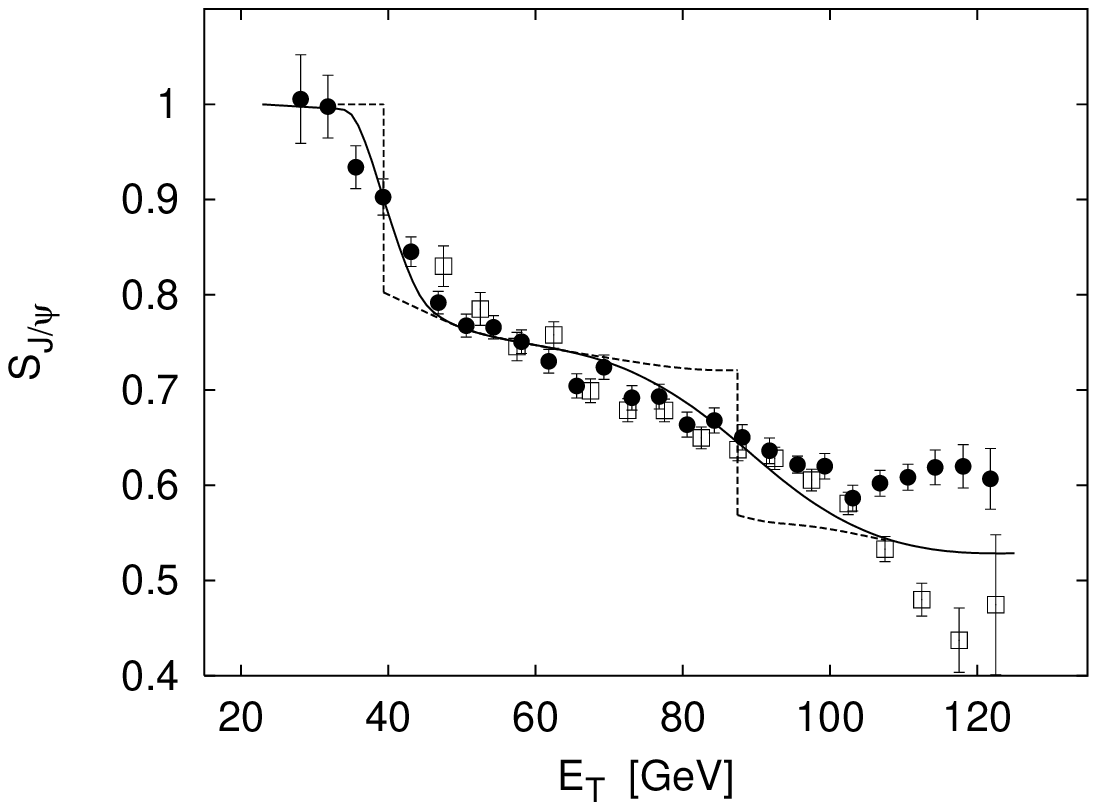,height= 80mm}}
\vspace*{-1.2cm}
\caption{The \J~survival probability as function of the measured transverse
energy $E_T$ for $Pb\!-\!Pb$ collisions at SPS energy; the dashed line
corresponds to collisions with fixed $E_T-b$ correlation, the solid line
includes the experimental $E_T-b$ smearing. The circles show the 1997,
the squares the 1998 data.}
\label{fig29}
\vspace*{-0.5cm}
\end{figure}

Schematically, one thus obtains a suppression pattern as illustrated in Fig.\ 
\ref{fig26}, where it is assumed that 40\% of the observed \J's come from
\X~decay and 60\% are produced directly. For clarity purposes, we
ignore here the small fraction (about 8\%) coming from \P~decay. The
dotted curve corresponds to the suppression which would take place in a
uniform medium of precisely specified parton density: at the
deconfinement point, all \X's are dissociated, so that the corresponding
fraction of decay \J's is gone; when the density for direct \J~melting
is reached, these disappear as well, leading to complete \J~suppression.
In actual nuclear collisions, the medium is not uniform, with denser
`inner' and less dense `outer' regions in the transverse plane.
Suppression now occurs only in the fraction $S_{\x}/S_{\rm total}$
for the \X~part and $S_{\j}/S_{\rm total}$ for the direct \J~part, where
$S_{\x}$ and $S_{\j}$ denote the clusters in which the respective
dissociation density is reached. Since these fractions depend on energy
and nuclear geometry, the suppression curves are different for
different experimental configurations. In Fig.\ \ref{fig26}, the 
result is shown
for $Pb\!-\!Pb$ collisions at the SPS, from impact parameter $b=11$ fm
to $b=0$ fm. The difference between this curve and the one for a uniform
medium thus reflects the surviving \J's produced in the less dense outer
regions.

In Fig.\ \ref{fig27}, similar calculations are shown for $Pb\!-\!Pb$ and
$Cu\!-\!Cu$ at RHIC energy. Here we note in particular that
$Pb\!-\!Pb$ collisions are for all meaningful centralities ($b \leq 11$
fm) above both the \X~and the direct \J~threshold, so that we get a
smooth anomalous suppression increasing from about 60\% at $b=11$ fm
to about 90\% at $b=0$. Combining this suppression with the `normal'
pre-resonance absorption in nuclear matter, we thus predict for central
$Pb\!-\!Pb$ collisions at RHIC a \J~production rate of less than 5\% the
corresponding unsuppressed rate (excluding possible $B$ decay
contributions). It should also be noted that this result is based on
twice the number of gluons per wounded nucleon at RHIC, compared to the
SPS value. A larger increase, based on a possible larger
hadron multiplicity at RHIC, would lead to more \J~suppression.
Similarly, we assume an average number $(dN_g/dy)_{y=0}$ of gluons
per wounded nucleon. It is conceivable that very central collisions
reach into the tail part of the multiplicity distribution, with a larger
number of hadrons and hence also gluons. This would lead to a larger
suppression for very central collisions. At the SPS, the NA50
collaboration can check if the basis for this exists. By combining
measurements of hadron multiplicity, transverse energy $E_T$ and forward
energy $E_{\rm ZDC}$, it is possible to study the number of hadrons per
wounded nucleon as function of $E_T$ and check if there is an increase
at highest $E_T$ values.

Finally we address briefly the comparison of our results on parton
percolation to the $E_T$-dependence of the actual data, which contain
an additional smearing due to the fact that a given $E_T$ bin
corresponds to a range of impact parameters and hence parton densities.
Including this effect in the standard way \cite{KLNS}, we obtain the
result shown in Fig.\ \ref{fig29}. Included in this figure are the 1997
\cite{NA50/97} and the 1998 data \cite{Cicalo,Kluberg} with minimum bias
determined Drell-Yan reference. The high $E_T$ points of the 1997 data
show an enhancement due to rescattering effects, which are removed in
the 1998 data using a thinner target. While reproducing the overall
behaviour, our results clearly show deviations from the data in detail. As
mentioned, the levelling-off of our curve at high $E_T$ would have to be
modified if an increase in the multiplicity per wounded nucleon
at high $E_T$ should be observed. Another possible modification could
enter through a density-dependent charmonium dissociation. We have here
assumed that all \X's melt once the critical density is reached.
Allowing a partial survival chance, which decreases with increasing
density, would lead to a steeper drop of the suppression with $n_{\rm
parton}$ as well as with $E_T$. One possible source for such an effect
would be the finite life-time of the deconfining medium in actual
nuclear collision, as seen e.g.\ in the parton cascade model to be
discussed in the next section.

Summarizing this section, we note that parton percolation provides a
consistent framework to study the onset of deconfinement as well as the
onset of charmonium suppression as a signature of the transition. It is
clear that the compatibility of the parton basis with other, soft hadron
production processes has to be checked \cite{Pajares}.

\section{COLOUR SCREENING IN THE PARTON CASCADE MODEL}

Probing deconfinement in nuclear collisions means studying the early
partonic medium left behind when the Lorentz-contracted nuclei have
passed through each other -- a medium consisting of partons, of
strings, of flux tubes, or of whatever basis is chosen. Any model of
this type has to

\begin{figure}[p]
\vspace*{-5mm}
\centerline{\psfig{file=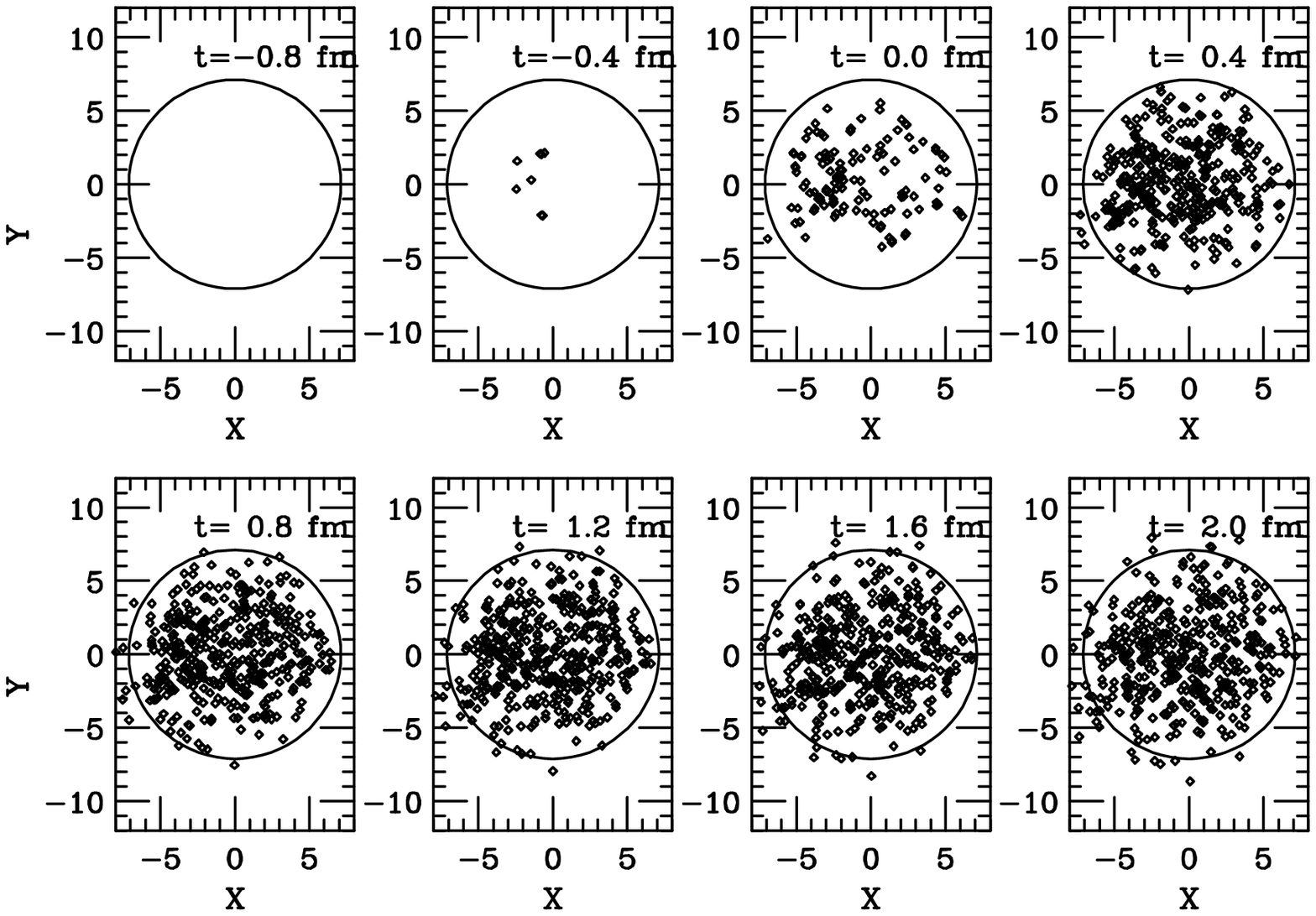,height= 100mm}}
\vspace*{-1cm}
\caption{Preliminary PCM calculations for
the time evolution of the parton population in the
transverse plane for central $Pb\!-\!Pb$ collisions at the SPS.}
\label{fig31b}
\end{figure}

\begin{figure}[p]
\vspace*{-0mm}
\centerline{\psfig{file=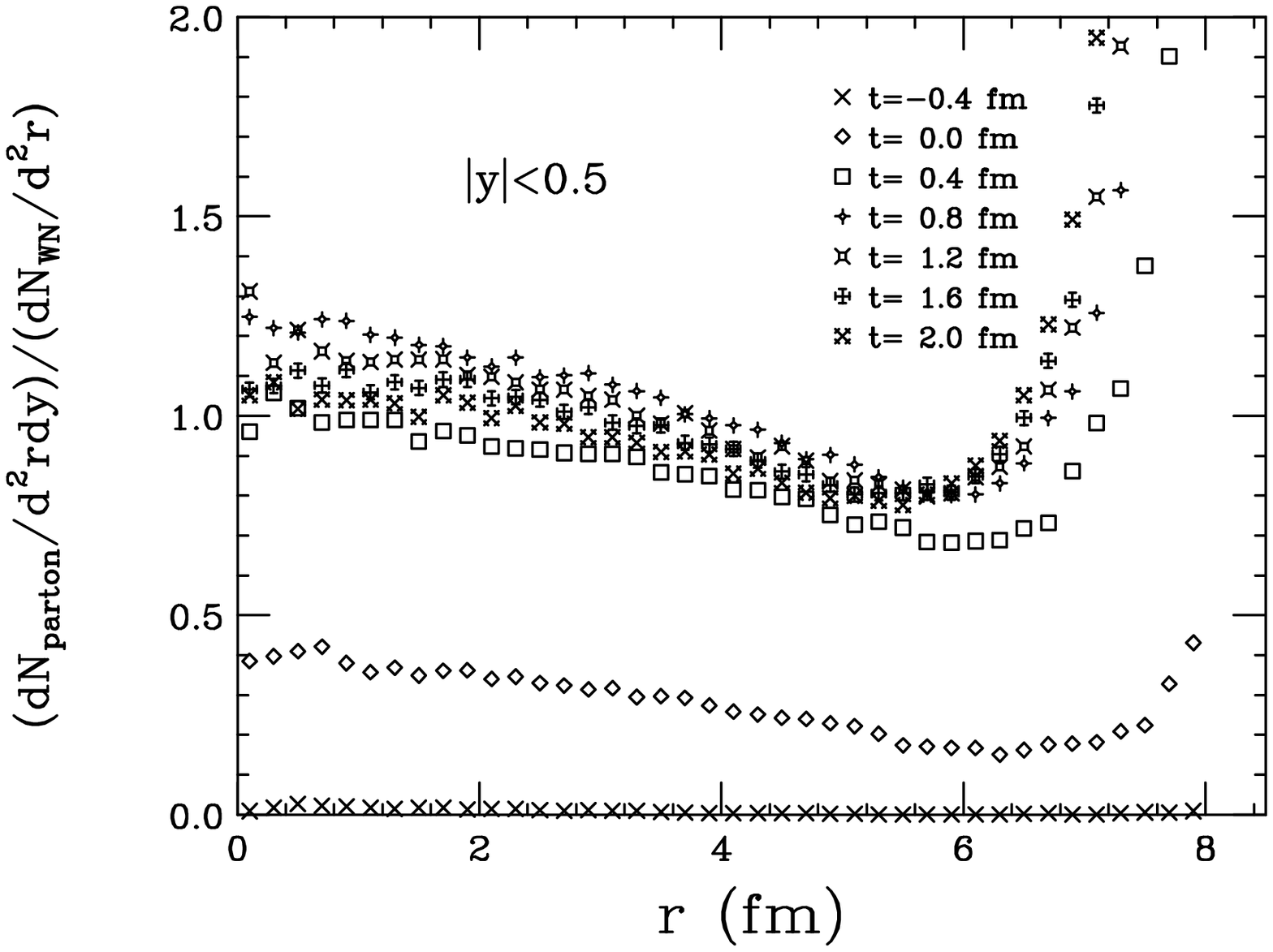,height= 100mm}}
\vspace*{-1cm}
\caption{Preliminary PCM calculations for 
the radial density of the parton population in the
transverse plane of central SPS $Pb\!-\!Pb$ collisions at different collision
times, compared to that obtained from the wounded nucleon distribution.}
\label{fig32}
\end{figure}

\begin{itemize}
\item{define deconfinement,}
\item{specify the onset of a particular signature such as charmonium
suppression, and}
\item{check that soft hadron production is consistent with the picture.}
\end{itemize}
In the past section, we had chosen a purely geometric framework, with
deconfinement defined as percolation of partons of transverse radius
$r \simeq 1/k_T$, and with a given (energy-dependent) number of
partons emitted from each wounded nucleon. In this section, we want to
consider a more detailed microscopic model, based on the interaction of
partons and the resulting evolution of the partonic medium \cite{Dinesh}.
For this, we shall use the parton cascade model \cite{Geiger/Mueller};
but evidently one could also use other approaches, such as HIJING
\cite{Wang} or the dual parton model \cite{Capella}. All such models
provide a parton or string distribution in the transverse plane and thus
in principle constitute an alternative to the nuclear profile pattern we
had used in the previous section. The time evolution of such a
distribution from the parton cascade model is shown in Fig.\ \ref{fig31b} and
compared to the nuclear profile pattern in Fig.\ \ref{fig32}. It is seen there
that the parton cascade model in fact provides a transverse parton
distribution very similar to that based on the distribution of wounded
nucleons.

The parton cascade model describes the partonic medium before the onset
of confinement. To determine when confinement sets in for such a medium,
we calculate the screening mass $\mu$ and recall that lattice QCD finds
$\mu_c \simeq 0.4 - 0.6$ GeV at the deconfinement point. The screening
mass is given by \cite{Biro}
\be
\mu^2 = -{3 \alpha_s \over \pi^2} \int d^3 k \nabla f(k), \label{3.1}
\ee
where $f(k)$ specifies the momentum distribution of the partons. For
a Bose-Einstein distribution, with $f(k)=(1-e^{-k/T})^{-1}$,
Eq.\ (\ref{3.1}) gives the familiar $\mu^2=4 \pi \alpha_s T^2$
obtained in perturbative QCD.

\begin{figure}[t]
\vspace*{-0mm}
\centerline{\psfig{file=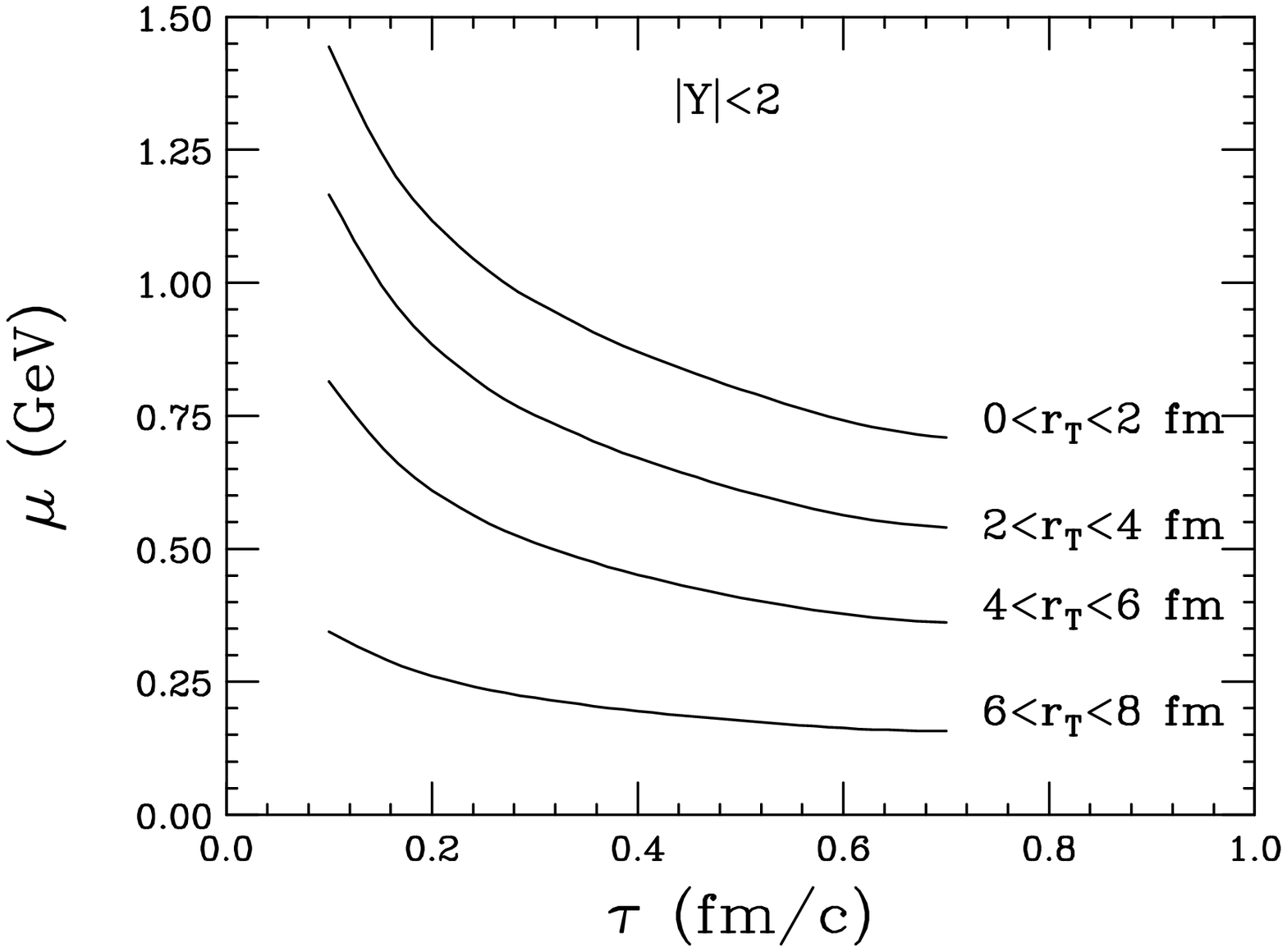,height= 80mm}}
\vspace*{-0.5cm}
\caption{Preliminary PCM calculations for
the time evolution of the colour screening mass in different
transverse regions of a central $Pb\!-\!Pb$ collision at RHIC energy.}
\label{fig33}
\end{figure}

Instead, we now insert for $f(k)$ the
time-dependent distribution function $f_{\rm PCM}(k,t)$ from the parton
cascade model, in order to obtain the non-equilibrium time evolution of
the screening mass. First preliminary results for central $Pb\!-\!Pb$
collisions at RHIC are given in Fig.\ \ref{fig33}, indicating 
the variation of $\mu$ as function of the proper medium
life-time after full nuclear overlap and as function of the transverse
radius of the medium. To interpret this behaviour,
recall that deconfinement as well as \X~dissociation occurs for $\mu
\geq$ 0.4 - 0.6 GeV, while the dissociation of directly produced \J's
requires 0.7 GeV. Hence for essentially all production regions, our
preliminary results for the screening mass exceed the necessary
dissociation values for both \X~and \J~for some time. In accord with the
results from parton percolaton, this would seem to indicate almost
complete \J~suppression for central $Pb\!-\!Pb$ collisions at RHIC.
To really conclude this, however, one must first specify how long $\mu$
has to exceed its critical value in order to `melt' the charmonium
state in question. In any case, for really quantitative precision
more extensive calculations are necessary.

\begin{figure}[h]
\vspace*{-0mm}
\centerline{\psfig{file=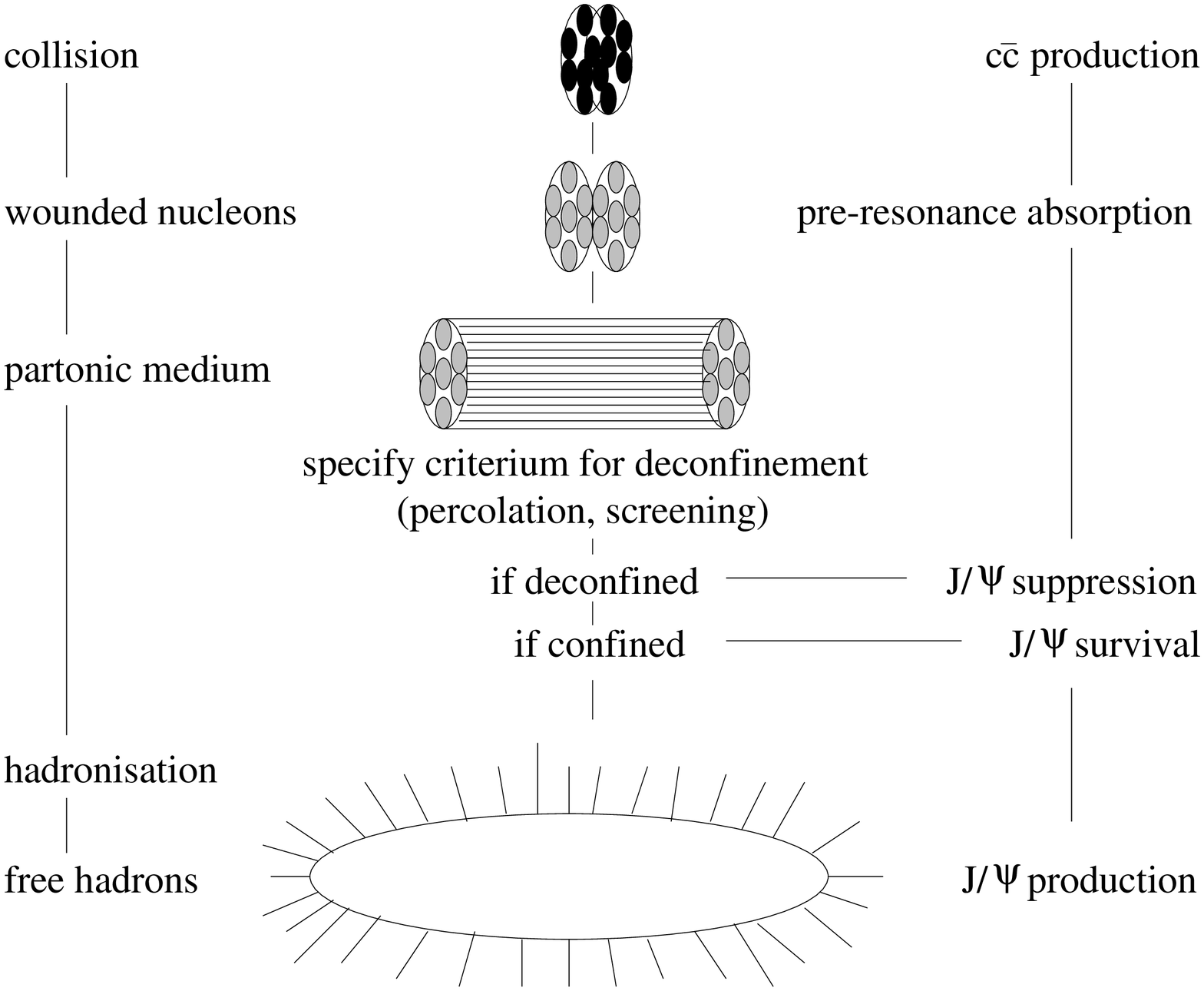,height= 100mm}}
\vspace*{0.5cm}
\caption{A short history of deconfinement}
\label{fig34}
\end{figure}

Let us finally note one particular advantage of microscopic evolution
picture such as the parton cascade model. Once we have defined the
conditions for quarkonium suppression, be it through percolation or
through colour screening, the parton cascade model can be used to check
if the conditions needed for the observed suppression lead to consistent
soft hadron production at the later times. Such models thus provide a
bridge between the early deconfined and the later hadronic stages.
The situation is summarized in Fig.\ \ref{fig34}, which also serves as
our conclusion. It shows on one hand the general evolution of the
collision, on the other the effect of this evolution on charmonium
production as hard probe of the early medium.

\bigskip

\leftline{\normalsize \bf ACKNOWLEDGEMENTS}

\medskip

It is a pleasure to thank Ph.\ Blanchard, S.\ Fortunato, H.-W.\
Huang, F.\ Karsch, D.\ Kharzeev, M.\ Nardi, D.\ K.\ Srivastava and D.\
Stauffer for helpful comments and discussions.


\begin{thebibliography}{99}
\bibitem{Cicalo} C.\ Cical\`o, these proceedings.
\bibitem{Kluberg} L.\ Kluberg, these proceedings.
\bibitem{MS} L.\ McLerran and B.\ Svetitsky, \PL 98 B (1981) 195
\bibitem{KPS} J.\ Kuti, J.\ Pol\'onyi and K.\ Szlach\'anyi, \PL 98 B
(1981) 199.
\bibitem{S-Y} B.\ Svetitsky and L.\ G.\ Yaffe, \NP B 210 [FS6] (1982)
423.
\bibitem{Engels} J.\ Engels et al., \PL B 365 (1996) 219
\bibitem{C-K} A.\ Coniglio and W. Klein, J.\ Phys.\ A 13 (1980) 2775.
\bibitem{S-A} For a recent survey, see D.\ Stauffer and A.\ Aharony,
{\sl Introduction to Percolation Theory}, Taylor \& Francis, London
1994.
\bibitem{FHS} S.\ Fortunato, H.-W.\ Huang and H.\ Satz, in preparation;
\par 
S.\ Fortunato, these proceedings.
\bibitem{KG} F.\ Green and F.\ Karsch, \NP B 238 (1984) 297.
\bibitem{Satz-P} H.\ Satz, \NP A642 (1998) 130c.
\bibitem{K} J.\ Kert\'esz, Physica A 161 (1989) 58.
\bibitem{Matsui} T.\ Matsui and H.\ Satz, \PL 178B (1986) 416.
\bibitem{KMS} F.\ Karsch, M.\ T.\ Mehr and H.\ Satz, \ZP C 37 (1988)
617.
\bibitem{KS} F.\ Karsch and H.\ Satz, \ZP C 51 (1991) 209.
\bibitem{Bialas} A.\ Bia\l as, M.\ Bleszy\'nski and W.\ Czyz, \NP B 111 (1976)
461. 
\bibitem{KNS} D.\ Kharzeev, M.\ Nardi and H.\ Satz, in preparation.
\bibitem{Pajares} N.\ Armesto et al., \PRL 77 (1996) 3736.
\bibitem{MRS-H} A.\ D.\ Martin, R.\ G.\ Roberts and W.\ J. Stirling,
Int.\ Journal of Mod.\ Phys.\ A 10 (1995) 2885. 
\bibitem{Alon} U.\ Alon, A.\ Drory and I.\ Balberg, \PR A 42 (1990)
4634.
\bibitem{W-S} C.\ W.\ deJager, H.\ deVries and C.\ deVries, Atomic Data
and Nuclear Data Tables 14 (1974) 485.
\bibitem{KLNS} D.\ Kharzeev, C.\ Louren{\c c}o, M.\ Nardi and H.\ Satz,
\ZP C 74 (1997) 307.
\bibitem{NS} M.\ Nardi and H.\ Satz, \PL B 442 (1998) 14.
\bibitem{NA50/97} {L.\ Ramello (NA50), \NP A 638 (1998) 261c \par
M.\ C.\ Abreu et al.(NA50), \PL B450 (1999) 456.}
\bibitem{Dinesh} D.\ K.\ Srivastava and H.\ Satz, in preparation
\bibitem{Geiger/Mueller} {K.\ Geiger and B.\ M\"uller, \NP B 369 (1992)
600\par
K.\ Geiger, Phys.\ Rep.\ 258 (1995) 376.}
\bibitem{Wang} M.\ Gyulassy and X.\-N.\ Wang, \PR D 45 (1992) 844.
\bibitem{Capella} See A.\ Capella et al., Phys.\ Rep.\ 236 (1994) 225.
\bibitem{Biro} T.\ S.\ Biro, B.\ M\"uller and X.-N.\ Wang, \PL B 283
(1992) 171.
\end{thebibliography}
\end{document}